\documentclass[preprint]{aastex}

\def\Msunh{~h^{-1}M_{\odot}\ }
\def\Mpc{~{\rm Mpc}}
\def\Mpch{h^{-1}{\rm Mpc}}
\def\kpch{h^{-1}{\rm kpc}}
\def\etal{{\it et al.}\ }

\def\vc#1{\mathbf{#1}}
\def\lesssim{{_ <\atop{^\sim}}}
\def\grtsim{{_ >\atop{^\sim}}}
\begin{document}

\title{Numerical Simulations in Cosmology I: Methods}
\author{Anatoly Klypin}
\affil{Astronomy Department, New Mexico State University, Box 30001, Department
4500, Las Cruces, NM 88003-0001}

%---------------------------------------------------------
\begin{abstract}
We give a short description of different methods used in cosmology.
The focus is on major features of $N$-body simulations: 
equations, main numerical techniques, effects of resolution, and
methods of halo identification.
\end{abstract}

%---------------------------------------------------------
\section{Introduction}
%---------------------------------------------------------

Numerical simulations in cosmology have a long history and numerous
important applications. Different aspects of the simulations including
history of the subject were reviewed recently by \citet{Bert}; see also 
\citet{Sellwood}. More detailed aspects of simulations 
were discussed by \citet{Gelb}, \citet{Gross}, and \citet{Kravtsov}.

Numerical simulations play a very significant role in cosmology. It all
started in 60s \citep{AarsethA} and 70s \citep{Peebles, PressSchecter}
with simple N-body problems solved using N-body codes with few hundred
particles. Later the Particle-Particle code
(direct summation of all two-body forces) was polished and brought to
the state-of-art \citep{AarsethB}. Already those early efforts brought
some very valuable fruits. \citet{Peebles} studied collapse of a cloud
of particles as a model of cluster formation. The model had 300 points
initially distributed within a sphere with no initial velocities. After
the collapse and virialization the system looked like a cluster of
galaxies. Those early simulations of cluster formation, though
producing cluster-like objects, signaled the first problem -- simple
model of initially isolated cloud (top-hat model) results in the
density profile of the cluster which is way too steep (power-law slope
-4) as compared with real clusters (slope -3). The problem was
addressed by \citet{GunnGott}, who introduced a notion of secondary
infall in an effort to solve the problem.  Another keystone work of
those times is the paper by \citet{WhiteA}, who studied collapse of 700
particles with different masses. It was shown that if one distributes
the mass of a cluster to individual galaxies, two-body
scattering will result in mass segregation not compatible with
observed clusters. This was another manifestation of the dark matter in
clusters. This time it was shown that inside a cluster the dark matter
can not reside inside individual galaxies.

Survival of substructures in galaxy clusters was another problem
addressed in the paper. It was found that lumps of dark matter, which
in real life may represent galaxies, do not survive in dense
environment of galaxy clusters. \citet{WhiteRees} argued that the real
galaxies survive inside clusters because of energy dissipation by the
baryonic component. That point of view was accepted for almost 20
years. Only recently it was shown the energy dissipation probably does
not play a dominant role in survival of galaxies and the dark matter
halos are not destroyed by tidal stripping and galaxy-galaxy collisions
inside clusters \citep{KGKK, Ghigna99}. The reason why early
simulations came to a wrong result was pure numerical: they did not
have enough resolution. But 20 years ago it was physically impossible
to make a simulation with sufficient resolution. Even if at that time
we had present-day codes, it would have taken about 600 years to make one
run.

Generation of initial condition with given amplitude and spectrum of
fluctuations was a problem for some time. The only correctly simulated
spectrum was the flat spectrum which was generated by randomly
distributing particles. In order to generate fluctuations with power
spectrum, say $P(k) \propto k^{-1}$, \citet{AarsethGT} placed particles
along rods. Formally, it generates the spectrum, but the distribution
has nothing to do with cosmological fluctuations.  \citet{Doroshkevich} and
\citet{KlypinShandarin} were the first to use the  \citet{Zeldovich}
approximation to set initial conditions. Since then this method is used to generate
initial conditions for arbitrary initial spectrum of perturbations.

Starting mid 80s the field of numerical simulations is blooming: new
numerical techniques are invented, old ones are perfected. The number
of publications based on numerical modeling skyrocketed.  To large
extend, this have changed our way of doing cosmology.  Instead of
questionable assumptions and waving-hands arguments, we have tools of
testing our hypothesis and models. As an example, I mention two
analytical approximations which were validated by numerical
simulations. The importance of both approximations is difficult to
overestimate. The first is the Zeldovich approximation, which paved the
way for understanding the large-scale structure of the galaxy
distribution. The second is the  \citet{PressSchecter}
approximation, which gives the number of objects formed at different
scales at different epochs. Both approximations cannot be formally
proved. The Zeldovich approximation formally is not applicable for
hierarchical clustering. It must start with smooth perturbations
(truncated spectrum). Nevertheless, numerical simulations have shown
that even for the hierarchical clustering the approximation can be used
with appropriate filtering of initial spectrum  
\citep[see][and references therein]{SahniColes}. The Press-Schechter
approximation is also difficult to justify without numerical
simulations. It operates with the initial spectrum and the linear theory, but
then (a very long jump) it predicts the number of objects
at very nonlinear stage. Because it is not based on any realistic
theory of nonlinear evolution, it was an
ingenious, but a wild guess. If anything, the approximation is based on
a simple spherical top-hat model. But simulations show that objects do
not form in this way -- they are formed in a complicated fashion
through multiple mergers and accretion along filaments. Still this a very
simple and a very useful prescription gives quite accurate
predictions.

This lecture is organized in the following way.  Section 2 gives the
equations which we solve to follow the evolution of initially small
fluctuations. Initial conditions are discussed in section 3.  A brief
discussion of different methods is given in section 4. Effects of the
resolution and some other technical details are also discussed in
Section 5. Identification of halos (``galaxies'') is discussed in
Section 6.

%---------------------------------------------------------
\section{Equations of evolution of fluctuations in an expanding
universe} 
%---------------------------------------------------------

Usually the problem of the formation and dynamics of cosmological
objects is formulated as $N$-body problem: for $N$ point-like objects
with given initial positions and velocities find their positions and
velocities at any later moment. It should be remembered that this just
a short-cut in our formulation -- to make things simple.  While it
still mathematically correct in many cases, it does not give a
correct explanation to what we do. If we are literally to take
this approach, we should follow the motion of zillions of axions,
baryons, neutrinos, and whatever else our Universe is made of. So, what
it has to do with the motion of those few millions of particles in our
simulations? The correct approach is to start with the Vlasov equation
coupled with the Poisson equation and with appropriate initial and
boundary conditions. If we neglect the baryonic component, which of
course is very interesting, but would complicate our situation even
more, the system is described by distribution functions $f_i({\bf x},
{\bf\dot x}, t)$ which should include all different clustered
components $i$. For a simple CDM model we have only one component
(axions or whatever it is). For more complicated Cold plus Hot Dark
Matter (CHDM) with few different types of neutrinos the system includes
one DF for the cold component and one DF for each type of neutrino
\citep{KHPR}. In the comoving coordinates {\bf x} the equations for
the evolution of $f_i$ are:

\begin{eqnarray}
\frac{\partial f_i}{\partial t} &+& 
     {\bf\dot x}\frac{\partial f_i}{\partial{\bf x}}
	\ - \ \nabla \phi\frac{\partial f_i}{\partial{\bf p}} = 0,
         \qquad {\bf p} = a^2 {\bf\dot x}, \\
  \nabla^2\phi &=& 4\pi G a^2 (\rho({\bf x},t)
           -\langle\rho_{\rm dm}(t)\rangle)
     \  = \ 4\pi G a^2\Omega_{\rm dm}\delta_{\rm dm}\rho_{\rm cr}, \\
\delta_{\rm dm}({\bf x}, t) &=& (\rho_{\rm dm}-\langle
        \rho_{\rm dm}\rangle)/\langle\rho_{\rm dm}\rangle), \\
\rho_{\rm dm}({\bf x}, t) &=&a^{-3}\sum_i m_i\int d^3{\bf p}
f_i({\bf x},{\bf\dot x},t).  
\label{eq:EqVlasov}
\end{eqnarray}

\noindent
Here $a=(1+z)^{-1}$ is the expansion parameter, ${\bf p}=a^2{\bf\dot x}$ is the
momentum, $\Omega_{\rm dm}$ is the contribution of the clustered dark
matter to the mean density of the Universe, $m_i$ is the mass of a
particle of $i-$th component of the dark matter. The solution of the
Vlasov equation can be written in terms of equations for characteristics,
which {\it look} like equations of particle motion:
\begin{eqnarray}
\frac{d{\bf p}}{da} &=& -\frac{\nabla\phi}{\dot a}, \qquad
   \frac{d{\bf v}}{dt}+2\frac{\dot a}{a}{\bf v} \ = \
-\frac{\nabla\phi\prime}{a^3} \label{eq:EqMotiona} \\
\frac{d{\bf x}}{da} &=& \phn \phn \frac{{\bf p}}{\dot a a^2}, \qquad	
\frac{d{\bf x}}{dt}  \ = \ {\bf v}\label{eq:EqMotionb} \\
\nabla^2\phi &=& 
        4\pi G\Omega_0\delta_{\rm dm}\rho_{\rm cr,0}/a, \quad 
 \phi\prime =a\phi\\
\dot a &=& H_0\sqrt{1+\Omega_0\left(\frac{1}{a}-1\right)
                  +\Omega_{\Lambda}\left(a^2-1\right)}
\label{eq:EqMotion}
\end{eqnarray}

\noindent
In these equations $\rho_{\rm cr,0}$ is the critical density at $z=0$;
$\Omega_0$, and $\Omega_{\Lambda,0}$, are the density of the matter and
of the cosmological constant in units of the critical density at $z=0$.

The distribution function $f_i$ is constant along each
characteristic. This property should be preserved by numerical
simulations. The complete set of characteristics coming through every
point in the phase space is equivalent to the
Vlasov equation. We can not have the complete (infinite) set, but we
can follow the evolution of the system (with some accuracy), if we
select a representative sample of characteristics. One way of doing
this would be to split initial phase space into small domains, take
only one characteristic as representative for each volume element, and
to follow the evolution of the system of the ``particles'' in
a self-consistent way. In models with one ``cold'' component of
clustering dark matter (like CDM or $\Lambda$CDM) the initial velocity
is a unique function of coordinates (only ``Zeldovich'' part is
present, no thermal velocities). This means that we need to split only
coordinate space, not velocity space. For complicated models with
significant thermal component, the distribution in full phase space
should be taken into account.  Depending on what we are interested in,
we might split initial space into equal-size boxes (typical setup for
PM or P$^3$M simulations) or we could divide some area of interest
(say, where a cluster will form) into smaller boxes, and use much
bigger boxes outside the area (to mimic gravitational forces of the
outside material). In any case, the mass assigned to a ``particle'' is
equal to the mass of the domain it represents. Now we can think of the
``particle'' either as a small box, which moves with the flow, but does
not change its original shape, or as a point-like particle. Both
presentations are used in simulations. None is superior to another.

There are different forms of final equations. Mathematically they are
all equivalent, but computationally there are very significant
differences. There are considerations, which may affect the choice of
particular form of the equations. Any numerical method gives more
accurate results for a variable, which changes slowly with time. For
example, for the gravitational potential we can choose either $\phi$ or
$\phi\prime$. At early stages of evolution perturbations still grow
almost linearly. In this case we expect that $\delta_{\rm dm}\propto
a$, $\phi\approx const$, and $\phi\prime\approx a$. Thus, $\phi$ can be a
better choice because it does not change. That is especially helpful, if
code uses gravitational potential from previous moment of time as
initial ``guess'' for current moment, as it happens in the case of the
ART code. In any case, it is better to have a variable, which does not
change much. For equations of motion we can choose, for example, either
first equations in eqs.(\ref{eq:EqMotiona}--~\ref{eq:EqMotionb}) or the second
equations.  If we choose ``momentum'' $p=a^2\dot x$ as effective
velocity and take the expansion parameter $a$
as time variable, then for the linear growth we expect that the change
of coordinates per each step is constant: $\Delta x\propto \Delta
a$. Numerical integration schemes should not have problem with this
type of growth. For the $t$ and $v$ variable, the rate of change is
more complicated:
$\Delta x\propto a^{-1/2}\Delta t$, which may produce some errors at
small expansion parameters. The choice of variables may affect
the accuracy of the solution even at very nonlinear stage of evolution
as was argued by \citet{Quinn}.

%---------------------------------------------------------
\section{Initial Conditions}
%---------------------------------------------------------

\subsection{Zeldovich approximation}

The Zeldovich approximation is commonly used to set initial conditions. The
approximation is valid in mildly nonlinear regime and is much 
superior to the linear approximation. We slightly rewrite the original
version of the approximation to incorporate cases (like CHDM) when the
growth rates $g(t)$ depend on the wavelength of the perturbation $|\vc{k}|$. In the
Zeldovich approximation the comoving and the lagrangian coordinates are
related in the following way:

\begin{equation}
\vc{x} =\vc{q} -\alpha\sum_{\vc{k}}g_{|\vc{k}|}(t)\vc{S}_{|\vc{k}|}(\vc{q}),
\quad 
\vc{p} = -\alpha a^2\sum_{\vc{k}}g_{|\vc{k}|}(t)
        \left(\frac {\dot g_{|\vc{k}|}}{ g_{|\vc{k}|}}\right)\vc{S}_{|\vc{k}|}(\vc{q}),
\label{eq:EqZeldtwo}
\end{equation}

\noindent
where the displacement vector $\vc{S}$ is related to the velocity
potential $\Phi$ and the power spectrum of fluctuations $P(|\vc{k}|)$:

\begin{equation}
\vc{S}_{|\vc{k}|}(\vc{q}) =\nabla_q\Phi_{|\vc{k}|}(\vc{q}), \quad 
\Phi_{|\vc{k}|}=\sum_{\vc{k}}a_{\vc{k}}\cos(\vc{k}\vc{q}) +
b_{\vc{k}}\sin(\vc{k}\vc{q}),
\label{eq:EqZeldthree}
\end{equation}

\noindent
where $a$ and $b$ are gaussian random numbers with the mean zero and
dispersion $\sigma^2=P(k)/k^4$:

\begin{equation}
a_{\vc{k}}=\sqrt{P(|\vc{k}|)}\cdot \frac{Gauss(0,1)}{ |\vc{k}|^2}, \quad 
b_{\vc{k}}=\sqrt{P(|\vc{k}|)}\cdot \frac{Gauss(0,1)}{ |\vc{k}|^2}.
\label{eq:EqZeldfour}
\end{equation}

The parameter $\alpha$, together with the power spectrum $P(k)$, define
the normalization of the fluctuations.

In order to set the initial conditions, we choose the size of the
computational box $L$ and the number of particles $N^3$. The phase
space is divided into small equal cubes of size $2\pi/L$. Each cube is
centered on a harmonic $\vc{k}=2\pi/L\times\{i,j,k\}$, where
$\{i,j,k\}$ are integer numbers with limits from zero to $N/2$. We make
a realization of the spectrum of perturbations $a_{\vc{k}}$ and
$b_{\vc{k}}$, and find displacement and momenta of particles with $\vc{q}=L/N\times\{i,j,k\}$ using  
eq.(\ref{eq:EqZeldtwo}). Here $i,j,k =1,N$.

\subsection{Power Spectrum}
There are approximations of the power spectrum $P(k)$ for a wide range of cosmological
models. Publicly available COSMICS code (Bertschinger 1996) gives
accurate approximations for the power spectrum.
Here we follow \citet{KlypinHoltzman} who give  the following fitting formula:

\begin{equation}
P(k) = \frac{k^n}{ (1 + P_2k^{1/2} +P_3k +P_4k^{3/2}+P_5k^{2})^{2P_6}}.
\label{eq:EqFit}
\end{equation}

\noindent 
The coefficients $P_i$ are presented by \citet{KlypinHoltzman} for a
variety of models.  The comparison of some of
the power spectra with the results from COSMICS (Bertschinger 1996)
indicate that the errors of the fits are smaller than 5\%.
Table~\ref{tab:spectrum} gives parameters of the fits for some popular models.

\begin{deluxetable}{rrrrrrrr} 
\tablecolumns{8} 
\tablewidth{0pc} 
\tablecaption{Approximations of Power Spectra} 
\tablehead{ 
\colhead{$\Omega_0$}    &  \colhead{$\Omega_{\rm bar}$}  &  
\colhead{$h$}   &  \colhead{$P_2$} &  \colhead{$P_3$} &
\colhead{$P_4$} &  \colhead{$P_5$} & \colhead{$P_6$} }
\startdata 
 0.3 & 0.035 & 0.60 & -1.7550E+00 & 6.0379E+01 & 2.2603E+02 & 5.6423E+02 & 9.3801E-01 \\
 0.3 & 0.030 & 0.65 & -1.6481E+00 & 5.3669E+01 & 1.6171E+02 & 4.1616E+02 & 9.3493E-01 \\
 0.3 & 0.026 & 0.70 & -1.5598E+00 & 4.7986E+01 & 1.1777E+02 & 3.2192E+02 & 9.3030E-01 \\
 1.0 & 0.050 & 0.50 & -1.1420E+00 & 2.9507E+01 & 4.1674E+01 & 1.1704E+02 & 9.2110E-01 \\
 1.0 & 0.100 & 0.50 & -1.3275E+00 & 3.0152E+01 & 5.5515E+01 & 1.2193E+02 & 9.2847E-01 \\
\enddata \label{tab:spectrum}
\end{deluxetable} 

The power spectrum of cosmological models is often approximated using
a fitting formula given by Bardeen \etal (1986, BBKS):

\begin{equation}
P(k)=k^nT^2(k), \quad T(k)={\ln(1+2.34q)\over 2.34q}
     [1+3.89q+(16.1q)^2+(5.4q)^3+(6.71q)^4]^{-1/4},
\label{eq:EqBBKS}
\end{equation}
\noindent 
where $q=k/(\Omega_0h^2\Mpc^{-1})$. Unfortunately, the accuracy of this
approximation is not great and it should not be used for accurate
simulations.  We find that the following approximation, which is a
combination of a slightly modified BBKS fit and the Hu \& Sugiyama
(1996) scaling with the amount of baryons, provides errors in the power
spectrum smaller than 5\% for the range of wavenumbers $k= (10^{-4} -
40)h\Mpc^{-1}$ and for $\Omega_b/\Omega_0<0.1$:

\begin{eqnarray}
P(k)  &=&k^nT^2(k), \nonumber\\
 T(k) &=&{\ln(1+2.34q)\over 2.34q}
     [1+13q+(10.5q)^2+(10.4q)^3+(6.51q)^4]^{-1/4},\nonumber\\
 q    &=&{k(T_{\rm CMB}/2.7K)^2\over
        \Omega_0h^2\alpha^{1/2}(1-\Omega_b/\Omega_0)^{0.60}},\qquad
 \alpha = a_1^{-\Omega_b/\Omega_0}a_2^{-(\Omega_b/\Omega_0)^3}\nonumber\\
 a_1 &=& (46.9\Omega_0h^2)^{0.670}[1+(32.1\Omega_0h^2)^{-0.532}],\quad
 a_2 = (12\Omega_0h^2)^{0.424}[1+(45\Omega_0h^2)^{-0.582}]
\label{eq:EqUgly}
\end{eqnarray}

\subsection{Multiple masses: high resolution for a small region}

In many cases we would like to set initial conditions in such a way
that inside some specific region(s) there are more particles and the
spectrum is better resolved. We need this when we want to have high
resolution for a halo, but we also need the environment of the halo.
This is done in a two-step process. First, we run a low resolution
simulation which has a sufficiently large volume to include the effects
of the environment. For this run all the particles have the same
mass. A halo is picked for rerunning with high resolution. Second,
using particles of the halo, we identify region in the lagrangian
(initial) space, where the resolution should be increased. We add
high-frequency harmonics, which are not present in the low resolution
run. We then add contributions of all the harmonics and get initial
displacements and momenta (eq.~\ref{eq:EqZeldtwo}). Let's be more
specific. In order to add the new harmonics, we must specify (1) how we
divide the phase space and place the harmonics, and (2) how we sum the
contributions of the harmonics.

\begin{figure}[tb!] 
%\epsscale{1.05} 
\plottwo{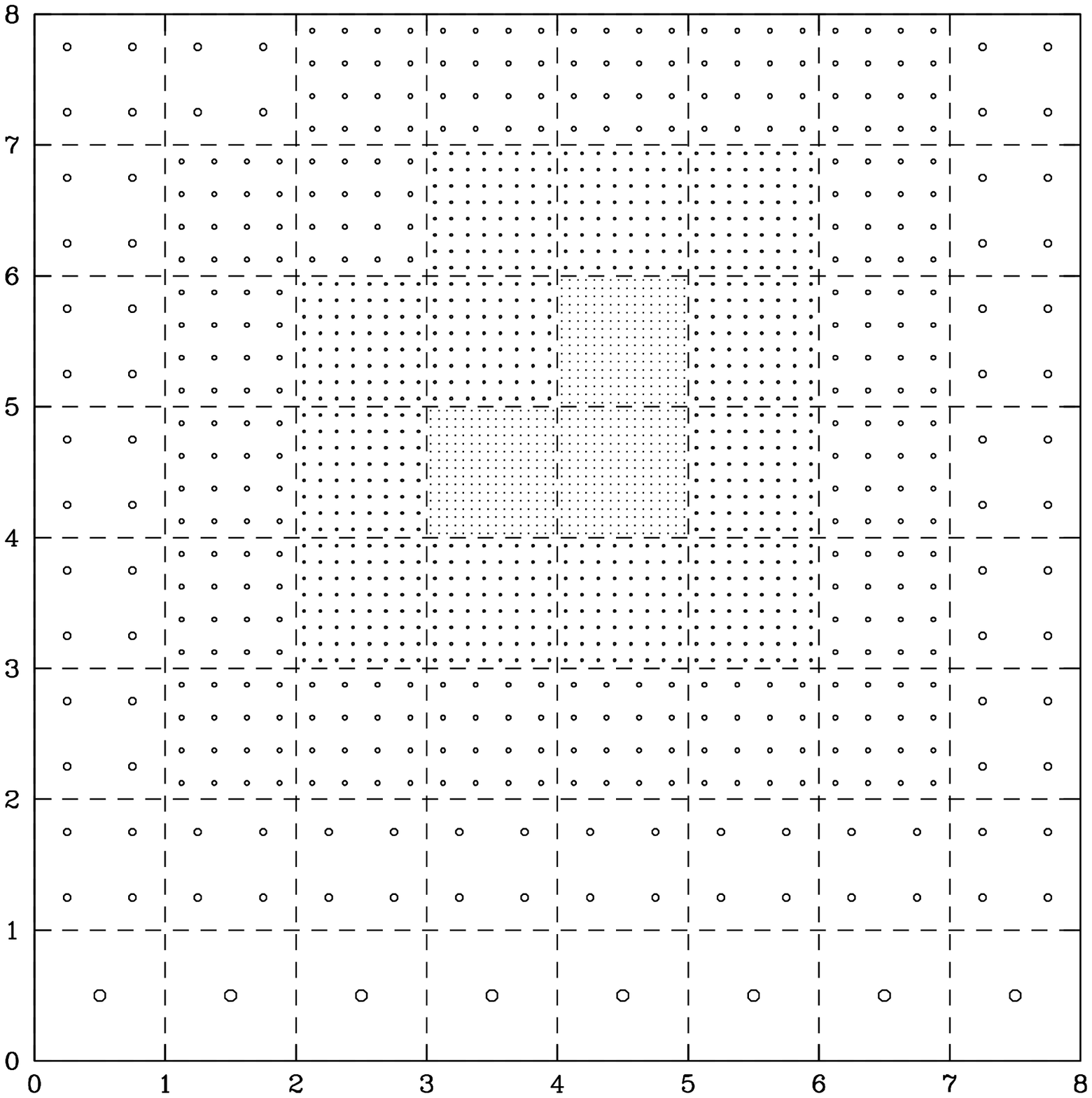}{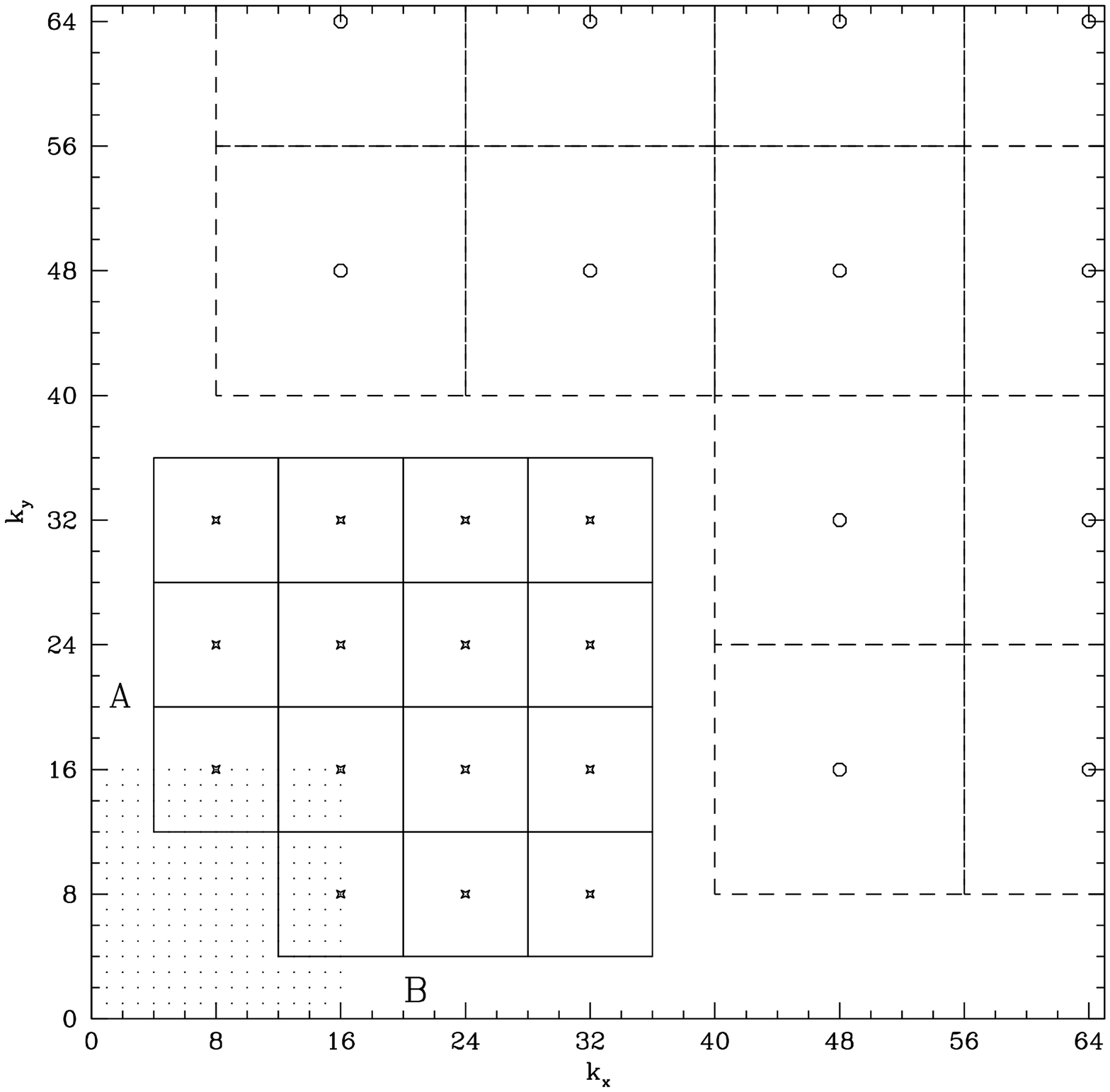}  
\caption{\small Example of construction of mass refinement in real
 space (left) and in phase space (right). In real space ({\it left
 panel}) three central blocks of particles were marked for highest mass
 resolution. Each block produces $16^2$ of smallest particles. Adjacent
 blocks get one step lower resolution and produce $8^2$ particles
 each. The procedure is repeated recursively producing fewer more
massive particles at each level. In phase space ({\it
 right panel}) small points in the left bottom corner represent
 harmonics used for the low resolution simulation.  For the high
 resolution run with box ratios 1:(1/8):(1/16) the phase space is sampled
 more coarsely, but high frequencies are included. Each harmonic
 (different markers) represents a small cube of the phase space
 indicated by squares. In this case the matching of the harmonics is
 not perfect: there are overlapping blocks and gaps. In any case, the waves
 inside domains A and B are missed in the simulation. }
\label{fig:gridexample}
\end{figure}

The simplest way  is to divide the phase space into many small
boxes of size $2\pi/L$, where $L$ is the box size. This is the same
devision, which we use to set the low resolution run. But now we extend
it to very high frequencies up to  $2\pi/L\times N/2$, where $N$ is the
new effective number of particles. For example, we used $N=64$ for the
low resolution run. For high resolution run we may choose
$N=1024$. Simply replace the value and run the code again. Of course,
we really can not do it because it would generate too many
particles. Instead, in some regions, where the resolution should not be
high, we combine particles together (by taking average coordinates and
average velocities) and replace many small-mass particles with fewer
larger ones. Left panel in Figure~\ref{fig:gridexample} gives an
example of mass refinement. Note that we try to avoid too large jumps
in the mass resolution by creating layers of particles of increasing
mass. 

\begin{figure}[tb!] 
\epsscale{0.5} 
\plotone{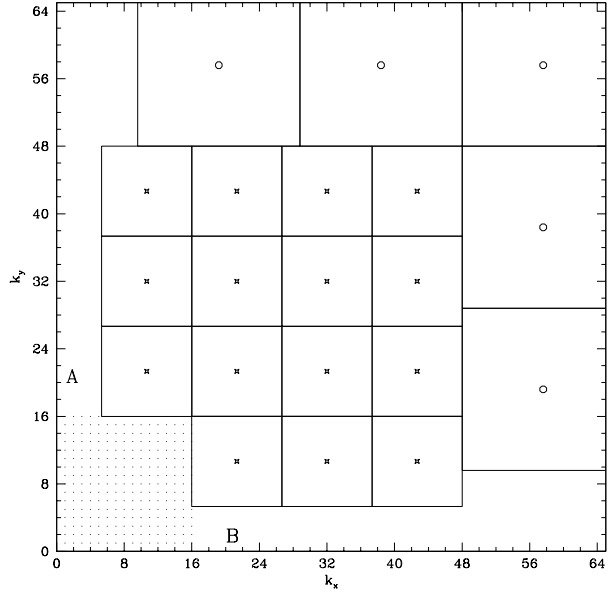}  
\caption{\small Another example of construction of mass refinement  in phase
space.  For the high resolution run with box ratios 1:(3/32):(5/96) the
phase space is sampled without overlapping blocks or gaps. }
\label{fig:gridgood}
\end{figure}

\begin{figure}[tb!] 
\epsscale{1.05} 
\plottwo{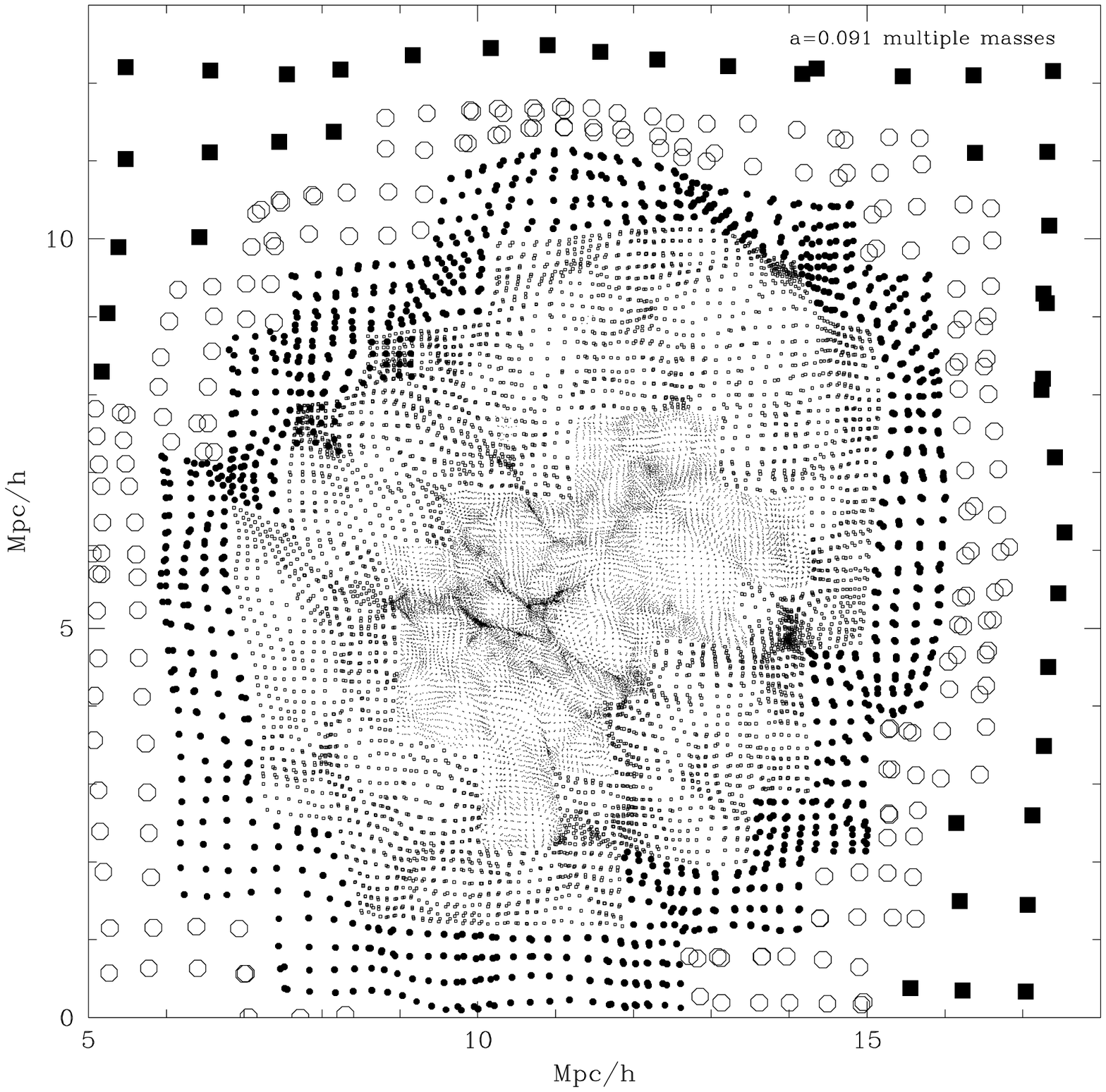}{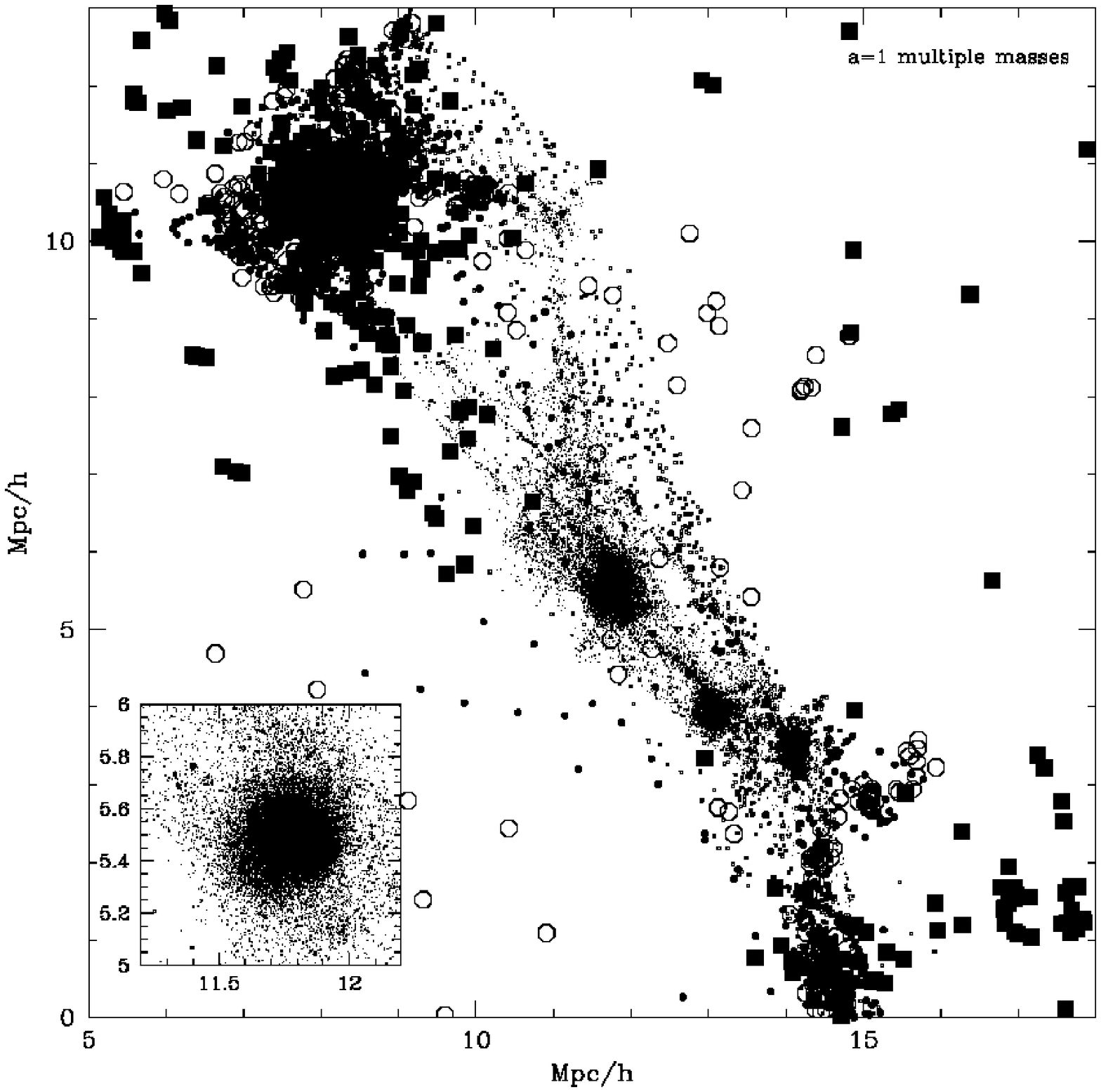} 
\caption{\small Distribution of particles of different masses in a thin 
slice going through the center of halo $A_1$ at redshift 10 (top 
panel) and at redshift zero (bottom panel). To avoid crowding of points 
the thickness of the slice is made smaller in the center (about $30\kpch$) 
and larger ($1\Mpch$) in the outer parts of the forming halo. Particles 
of different mass are shown with different symbols. }   
\label{fig:Massexample} 
\end{figure}

This approach is correct and relatively simple. It may seem that it
takes too much cpu to get the initial conditions. In practice, cpu time
is not much of an issue because initial conditions are generated only
once and it takes only few cpu hours even for $1024^3$ mesh. For most
of applications $1024^3$ particles is more then enough. The problem
arises when we want to have more then $1024^3$ particles. We simply do
not have enough computer memory to store the information for all the
harmonics. In this case we must decrease the resolution in the phase
space. It is a bit easier to understand the procedure, if we consider
phase space diagrams like one presented in
Figure~\ref{fig:Massexample}. The low resolution run in this case was
done for $32^3$ particles with harmonics up to $16\times2\pi/L$ (small
points). For the high resolution run we choose a region of size 1/8 of
the original large box. Inside the small box we place another box,
which is twice smaller. Thus, we will have three levels of mass
refinement. For each level we have corresponding size of the phase
space block. The size is defined by the size of real space box and is
equal to $2\pi/L\times K$, $K=1, 8, 16$. Harmonics from different
refinements should not overlap: if a region in phase space is
represented on lower level of resolution, it should not appear in the
the higher resolution level. This is why rows of the highest resolution
harmonics (circles) with $K_x=16$ and $K_y=16$ are absent in the
Figure~\ref{fig:Massexample}: they have been already covered by lower
resolution blocks marked by stars. Figure~\ref{fig:Massexample} clearly
illustrate that matching harmonics is a complicated process: we failed
to do the match because there are partially overlapping blocks and
there are gaps. We can get much better results, if we assume different
ratios of the sizes of the boxes. For example, if instead of box ratios
$1:(1/8):(1/16)$, we chose ratios $1:(3/32):(5/96)$, the coverage of the phase
space is almost perfect as shown in Figure:~\ref{fig:gridgood}.

\section{Codes}
There are many different numerical techniques to follow the evolution
of a system of many particles. For earlier reviews see
\citet{HockneyEast, Sellwood}, and \citet{Bert}. Most of the
methods for cosmological applications take some ideas from three
techniques: Particle Mesh (PM) code, direct summation or
Particle-Particle code, and the TREE code. For example, the Adaptive
Particle-Particle/Particle-Mesh (AP$^3$M) code \citep{Couchman}
is a combination of the PM code and the Particle-Particle code. The
Adaptive-Refinement-Tree code (ART) \citep{ART, Kravtsov} is an extension of 
the PM code with the organization of meshes in the form of a tree.
All methods have their advantages and disadvantages.

{\bf PM code}. It uses a mesh to produce density and potential.  As
the result, its resolution is limited by the size of the mesh.
There are two advantages of the method: i) it is fast (the smallest
number of operations per particle per time step of all the other
methods), ii) it typically uses very large number of particles. The
later can be crucial for some applications. There are few
modifications of the code. 
``Plain-vanilla'' PM was described by \citep{HockneyEast}. 
It includes Cloud-In-Cell density assignment and
7-point discrete analog of the Laplacian operator. Higher order
approximations improve the accuracy on large distances, but degrades
the resolution \citep[e.g.,][]{Gelb}. The PM code is available \citep{KlypinHoltzman}

{\bf P$^3$M code} is described in detail in \citet{HockneyEast} and
\citet{Efff}. It has two parts: PM part, which
takes care of large-scale forces, and PP part, which adds small-scale
particle-particle contribution. Because of strong
clustering at late stages of evolution, PP part becomes prohibitively
expensive once large objects start to form in large
numbers. Significant speed is achieved in modified version of the
code, which introduces sub-grids (next levels of PM) in areas with high
density \citep{Couchman}. With modification the code is as fast as the TREE
code even for heavily clustered configurations.
The code express the inter-particle force as a sum of
a short range force (computed by direct particle-particle pair force
summation) and the smoothly varying part (approximated by the
particle-mesh force calculation). One of the major problems for these
codes is the correct splitting of the force into a short-range and a
long-range part. The grid method (PM) is only able to produce reliable
inter particle forces down to a minimum of at least two grid cells.
For smaller separations the force can no longer be represented on the
grid and therefore one must introduce a cut-off radius $r_e$ (larger
than two grid cells), where for $r < r_e$ the force should smoothly
go to zero.  The parameter $r_e$ defines the chaining-mesh and for 
distances smaller than this cutoff radius $r_e$ a
contribution from the direct particle-particle (PP) summation needs to
be added to the total force acting on each particle. Again this PP
force should smoothly go to zero for very small distances in order to
avoid unphysical particle-particle scattering.  This cutoff of the PP
force determines the overall force resolution of a P$^3$M code.

The most widely used version of this algorithm is currently the
adaptive P$^3$M (AP$^3$M) code of Couchman (1991), which is available
for public.. The smoothing of the force in this code is connected to a
$S_2$ sphere, as described in Hockney \& Eastwood (1981).

{\bf TREE code} is the most flexible code in the sense of the choice of
boundary conditions \citep{Appel, BarnesHut,Hernquist}. It is also more
expensive than PM: it takes 10-50 times more operations.
\citep{BouchetHernquist} and \citep{HSB} extended the code for the
periodical boundary conditions, which is important for simulating
large-scale fluctuations. Some variants of the TREE are publicly
available. There are variants of the code modified for massively
parallel computers. There code variants with variable time
stepping, which is vital for extremely high resolution simulations.

{\bf ART code}. Multigrid methods were introduced long ago, but only
recently they started to show a potential to produce real results.  It
worth of paying attention if a ``multigrid'' code is really a fully
adaptive multigrid code. An example of this type of the codes is the
Adaptive Refinement Tree code (ART; Kravtsov et al. 1997), which
reaches high force resolution by refining all high-density regions with
an automated refinement algorithm.  The refinements are recursive: the
refined regions can also be refined, each subsequent refinement having
half of the previous level's cell size.  This creates an hierarchy of
refinement meshes of different resolutions covering regions of
interest.  The refinement is done cell-by-cell (individual cells can be
refined or de-refined) and meshes are not constrained to have a
rectangular (or any other) shape. This allows the code to refine the
required regions in an efficient manner.  The criterion for refinement
is the {\em local overdensity} of particles the code refines an individual cell only if the
density of particles (smoothed with the cloud-in-cell scheme; Hockney
\& Eastwood 1981) is higher than $n_{th}$ particles, with typical
values $n_{th}=2-5$.  The Poisson equation on the hierarchy of meshes
is solved first on the base grid using FFT technique and then on the
subsequent refinement levels. On each refinement level the code obtains
the potential by solving the Dirichlet boundary problem with boundary
conditions provided by the already existing solution at the previous
level or from the previous moment of time. There is no
particle-particle summation in the ART code and the actual force
resolution is equal to $\approx 2$ cells of the finest refinement mesh
covering a particular region.

\begin{figure}[tb!] 
\epsscale{0.5} 
\plotone{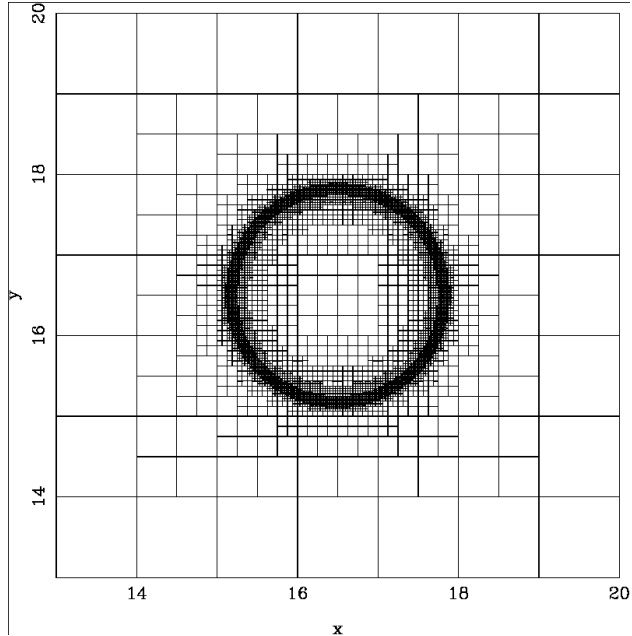}
\caption{\small  An example of a refinement structure
constructed by (hydro)ART code for spherical strong explosion  (courtesy of A. Kravtsov)  }   
\label{fig:ARTrefinement} 
\end{figure}

Figure~\ref{fig:ARTrefinement} (courtesy of A. Kravtsov) gives an
example of mesh refinement for hydro-dynamical version of the ART code.
The code produced this refinement mesh for spherical strong explosion
(Sedov solution).

The refinement of the time integration mimics spatial refinement and
the time step for each subsequent refinement level is two times smaller than
the step on the previous level. Note, however, that particles on the
same refinement level move with the same step. When a particle moves
from one level to another, the time step changes and its position and
velocity are interpolated to appropriate time moments. This interpolation
is first-order accurate in time, whereas the rest of the integration is done 
with the second-order accurate time centered leap-frog scheme. All equations
are integrated with the expansion factor $a$ as a time variable and the
global time step hierarchy is thus set by the step $\Delta a_0$ at the
zeroth level (uniform base grid). The step on level $L$ is then
$\Delta a_L=\Delta a_0/2^L$.  

What code is the best? Which one to choose?  There is no unique answer
-- everything depends on the problem, which we are addressing. For
example, if we are interested in explanation of the large-scale
structure (filaments, voids, Zeldovich approximation, and so on), PM
code with 256$^3$ mesh is sufficient. It takes only one night to make a
simulation on a (good) workstation. There is a very long list of
problems like that.  But if you intent to look for the structure of
individual galaxies in the large-scale environment, you must have a
code with much better resolution with variable time stepping, and with
multiple masses.  In this case the TREE or ART codes are the
choices.

%---------------------------------------------------------
\section{Effects of resolution}
%---------------------------------------------------------

As the resolution of simulations improves and the range of their
applications broaden, it becomes increasingly important to understand
the limits of the simulations.  
\citet{Knebe} made detailed comparison of realistic simulations done
with three codes: ART, AP$^3$M, and PM. Here we present some of their
results and main conclusions. The simulations were done for the
standard CDM model with the dimensionless Hubble constant $h=0.5$ and
$\Omega_0=1$. The simulation box of $15\Mpch$ had $64^3$ equal-mass
particles, which gives the mass resolution (mass per particle) of
$3.55\times 10^9\Msunh$. Because of the low resolution of the PM runs,
we show results only for the other two codes. For the ART code the
force resolution is practically fixed by the number of particles. The
only free parameter is the number of steps on the lowest (zero) level
of resolution. In the case of the AP$^3$M, besides the number of steps,
one can also request the force resolution. Parameters of two runs with
the ART code and five simulations with the AP$^3$M are given in
Table~\ref{tab:param}.
\begin{table}
\caption{Parameters of the numerical simulations.  }
\label{tab:param}
 \begin{center}
 \begin{tabular}{lcccc} \hline
{\bf simulation}& softening
                                               & dyn. range & steps &
$N_{\rm steps}$/dyn.range \\
 & ($h^{-1}$kpc)& & (min-max) & \\ \hline
 AP$^3$M$_1$ & 3.5 &    4267    & 8000 & 1.87 \\
 AP$^3$M$_2$ & 2.3 &    6400    & 6000 & 0.94 \\
 AP$^3$M$_3$ & 1.8 &    8544    & 6000 & 0.70 \\
 AP$^3$M$_4$ & 3.5 &    4267    & 2000 & 0.47 \\
 AP$^3$M$_5$ & 7.0 &    2133    & 8000 &  3.75\\
 ART$_1$  & 3.7 &    4096    & 660-21120 & 2.58   \\
 ART$_2$  & 3.7 &    4096    & 330-10560  & 5.16 \\ \hline
 \end{tabular}
 \end{center}
 \end{table}

\begin{figure}[tb!] 
	\epsscale{0.7} 
      \plotone{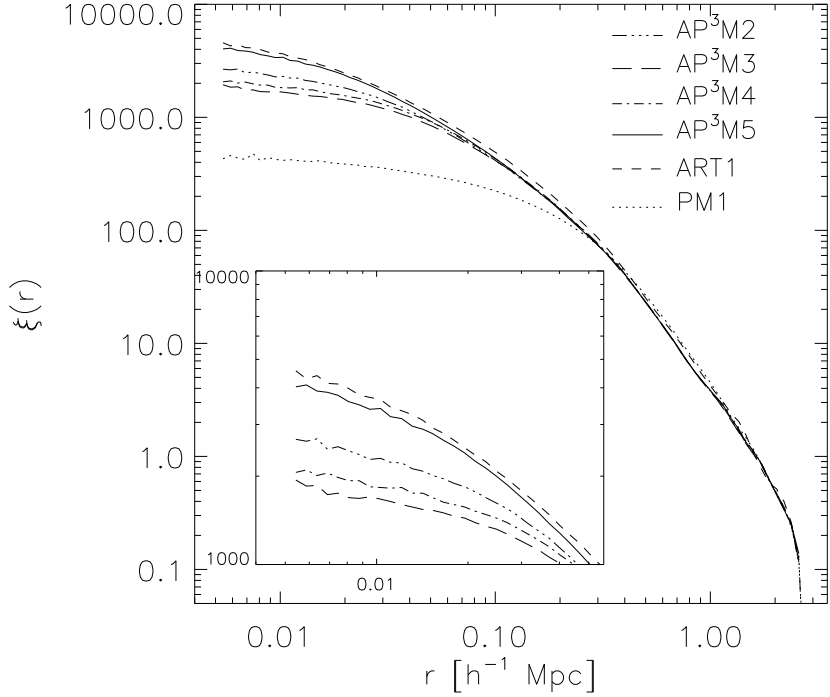}
      \caption{\small Correlation function of dark matter particles. Note that the range 
                of correlation amplitudes is different in the inset panel.}
      \label{fig:DMxi}
   \end{figure}

Figure~\ref{fig:DMxi} shows the correlation function for the dark matter
down to the scale of $5\kpch$, which is close to the force resolution
of all our high-resolution simulations. The correlation function in
runs AP$^3$M$_1$ and ART$_2$ are similar to those of AP$^3$M$_5$ and ART$_1$
respectively and are not shown for clarity. We can see that the
AP$^3$M$_5$ and the ART$_1$ runs agree to $\lesssim 10\%$ over the whole
range of scales. The correlation amplitudes of runs AP$^3$M$_{2-4}$,
however, are systematically lower at $r\lesssim 50-60\kpch$ (i.e., the
scale corresponding to $\approx 15-20$ resolutions), with the AP$^3$M$_3$
run exhibiting the lowest amplitude.  The fact that the AP$^3$M$_2$
correlation amplitude deviates less than that of the AP$^3$M$_3$ run,
indicates that the effect is very sensitive to the force resolution.

Note that the AP$^3$M$_3$ run has formally the best force
resolution. Thus, one would naively expect that it should gives the
largest correlation function. At scales $\lesssim 30\kpch$ the
deviations of the AP$^3$M$_3$ from the ART$_1$ or the AP$^3$M$_5$ runs are
$\approx 100-200\%$. We attribute these deviations to the numerical
effects: high force resolution in AP$^3$M$_3$ was not adequately supported
by the time integration. In other words, the AP$^3$M$_3$ had too few
time-steps. Note that it had a quite large number of steps (6000), not
much smaller than the AP$^3$M$_5$ (8000). But for its force resolution, it
should have many more steps. The lack of the number of steps was
devastating. 
 
   \begin{figure}[tb!] 
	\epsscale{0.7} 
     \plotone{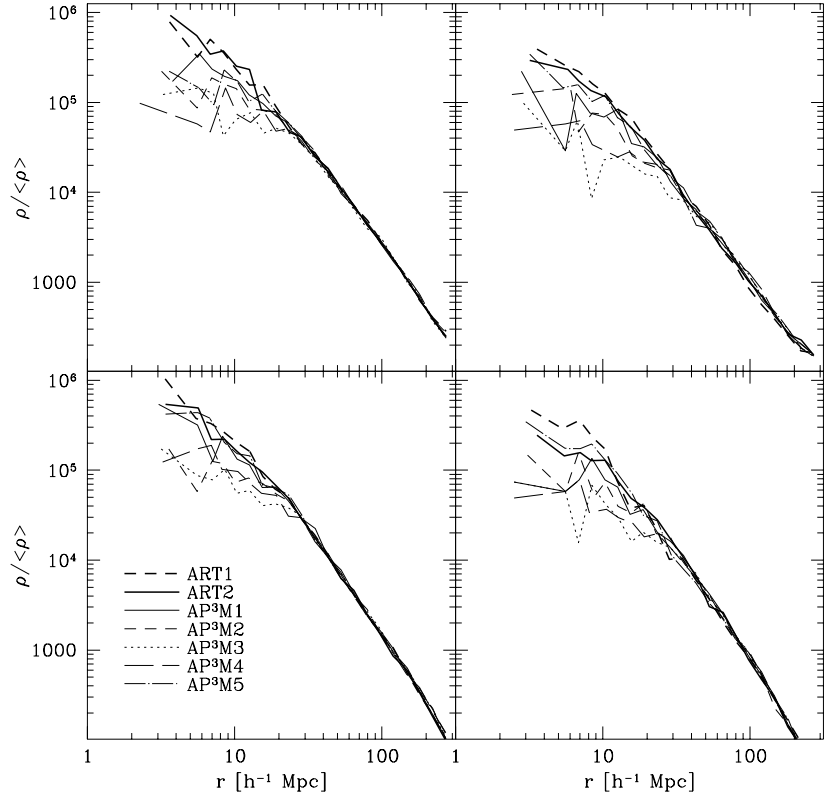} 
    \caption{\small Density profiles of four
   largest halos in simulations of \citet{Knebe}. Note that the AP$^3$M$_3$
   run has formally the best force resolution, but its actual
   resolution was much lower because of insufficient number of steps.}
   \label{fig:haloprofile} \end{figure}

Figure~\ref{fig:haloprofile} presents the density profiles of four of the most massive halos in our
simulations. We have not shown the profile of the most massive halo
because it appears to have undergone a recent major merger and is not
very relaxed. In this figure, we present only profiles of halos in the
high-resolution runs. Not surprisingly, the inner density of the PM
halos is much smaller than in the high-resolution runs and their
profiles deviate strongly from the profiles of high-resolution halos
at the scales shown in Figure~\ref{fig:haloprofile}. 
A glance at Figure~\ref{fig:haloprofile} shows that all profiles agree well at
$r\grtsim 30\kpch$. This scales is about eight times smaller than the
mean inter-particle separation. Thus, despite the very different
resolutions, time steps, and numerical techniques used for the
simulations, the convergence is observed at a scale much lower than
the mean inter-particle separation, argued by Splinter et al. (1998) to
be the smallest trustworthy scale.

Nevertheless, there are systematic differences between the runs.  The
profiles in two ART runs are identical within the errors indicating
convergence (we have run an additional simulation with time steps
twice smaller than those in the ART$_1$ finding no difference in the
density profiles). Among the AP$^3$M runs, the profiles of the AP$^3$M$_1$
and AP$^3$M$_5$ are closer to the density profiles of the ART halos than
the rest. The AP$^3$M$_2$, AP$^3$M$_3$, and AP$^3$M$_4$, despite the higher
force resolution, exhibit lower densities in the halo cores, the
AP$^3$M$_3$ and AP$^3$M$_4$ runs being the most deviant. 

These results can be interpreted, if we examine the trend of the
central density as a function of the ratio of the number of time steps
to the dynamic range of the simulations (see Table~\ref{tab:param}). The
ratio is smaller when either the number of steps is smaller or the
force resolution is higher.  The agreement
in density profiles is observed when this ratio is $\grtsim 2$. This
suggests that for a fixed number of time steps, there should be a limit
on the force resolution. Conversely, for a given force resolution,
there is a lower limit on the required number of time steps. The exact
requirements would probably depend on the code type and the integration
scheme.  For the AP$^3$M code our results suggest that the ratio of the
number of time steps to the dynamic range should be no less than
one. It is interesting that the deviations in the density profiles are
similar to and are observed at the same scales as the deviations in the
DM correlation function (Fig.~\ref{fig:DMxi}) suggesting that the
correlation function is sensitive to the central density distribution
of dark matter halos.

%---------------------------------------------------------
\section{Halo identification}
%---------------------------------------------------------
Finding halos in dense environments is a challenge. 
 Some of the problems that any halo finding algorithm
faces are not numerical. They exist in the real Universe. We select a few
typical difficult situations.

1. {\it A large galaxy with a small satellite.} Examples: LMC and the
Milky Way or the M51 system.  Assuming that the satellite is bound, do
we have to include the mass of the satellite in the mass of the large
galaxy? If we do, then we count the mass of the satellite twice: once when
we find the satellite and then when we find the large galaxy. This does
not seem reasonable. If we do not include the satellite, then the mass
of the large galaxy is underestimated. For example, the binding energy
of a particle at the distance of the satellite will be wrong. The
problem arises when we try to assign particles to different halos in
an effort to find masses of halos. This is very difficult to do for
particles moving between halos. Even if a particle at some moment has
negative energy relative to one of the halos, it is not guaranteed that
it belongs to the halo. The gravitational potential changes with time,
and the particle may end up falling onto another halo. This is not just
a precaution. This actually was found very often in real halos when we
compared contents of halos at different redshifts. Interacting halos
exchange mass and lose mass. We try to avoid the situation: instead of
assigning mass to halos, we find the maximum of the ``rotational
velocity'', $\sqrt{GM/R}$, which is observationally a more meaningful
quantity.

2. {\it A satellite of a large galaxy.} The previous situation is now
viewed from a different angle. How can we estimate the mass or the
rotational velocity of the satellite? The formal virial radius of the
satellite is large: the big galaxy is within the radius. The rotational
velocity may rise all the way to the center of the large galaxy. In
order to find the outer radius of the satellite, we analyze the density
profile. At small distances from the center of the satellite the
density steeply declines, but then it flattens out and may even
increase. This means that we reached the outer border of the
satellite. We use the radius at which the density starts to flatten out
as the first approximation for the radius of the halo. This
approximation can be improved by removing unbound particles and
checking the steepness of the density profile in the outer part.

3. {\it Tidal stripping.} Peripheral parts of galaxies, responsible for
extended flat rotation curves outside of clusters, are very likely 
tidally stripped and lost when the galaxies fall into a cluster. The
same happens with halos: a large fraction of halo mass may be lost due
to stripping in dense cluster environments. 
Thus, if an algorithm finds that 90\% of mass of a halo identified at
early epoch is lost, it does not mean that the halo was destroyed. This
is not a numerical effect and is not due to  ``lack of physics''. This
is a normal situation. What is left of the halo, given that it still has a large
enough mass and radius, is a ``galaxy''.

There are different methods of identifying collapsed 
objects (halos) in numerical simulations.  

{\bf Friends-Of-Friends (FOF)} algorithm was used a lot and still has
its adepts. If we imagine that each particle is surrounded by a sphere
of radius $b d/2$, then every connected group of particles is
identified as a halo. Here $d$ is the mean distance between particles,
and $b$ is called {\it linking parameter}, which typically is 0.2.
Dependence of groups on $b$ is extremely strong. The method stems from
an old idea to use percolation theory to discriminate between
cosmological models.  Because of that, FOF is also called percolation
method, which is wrong because the percolation is about groups
spanning the whole box, not collapsed and compact objects. FOF was
criticized for failing to find separate groups in cases when those
groups were obviously present
\citep{Gelb}. The problem originates from the tendency of FOF to 
``percolate'' through bridges connecting interacting galaxies or
galaxies in high density backgrounds. 

{\bf DENMAX} tried to overcome the problems of FOF by dealing with
density maxima \citep{Gelb, GelbBert}. It finds maxima of
density and then tries to identify particles, which belong to each
maximum (halo). The procedure is quite complicated. First, density
field is constructed. Second, the density (with negative sign) is
treated as potential in which particles start to move as in a viscous
fluid.  Eventually, particles sink at bottoms of the potential (which
are also maxima density). Third, only particles with negative energy
(relative to their group) are retained. Just as in the case of FOF, we
can easily imagine situations when (this time) DENMAX should fail. For
example, two colliding galaxies in a cluster of galaxies. Because of
large relative velocity they should just pass each other. In the
moment of collision DENMAX ceases to ``see'' both galaxies because all
particle have positive energies. That is probably a quite unlikely
situation. The method is definitely one of the best at present. The
only problem is that it seems to be too complicated for present state
of simulations. DENMAX has two siblings -- SKID (Stadel et al.) and BDM
\citep{KlypinHoltzman} -- which are frequently used.
 
{\bf ``Overdensity 200''}. There is no name for the method, but it is
often used. Find density maximum, place a sphere and find radius,
within which the sphere has the mean overdensity 200 (or 177 if you
really want to follow the top-hat model of nonlinear collapse).


\begin{thebibliography}{}
 
\bibitem[Aarseth(1963)]{AarsethA} Aarseth, S.J. 1963, \mnras, 126, 223
\bibitem[Aarseth et al.(1979)]{AarsethGT} Aarseth S.J., Gott J.R.,
	 \& Turner E.L. 1979, \apj, 228, 664
\bibitem[Aarseth (1985)]{AarsethB} Aarseth S.J. 1985, in {\it Multiple Time Scales} edited by
	J. W. Brackbill and B. J. Cohen (New York, Academic), p.377
\bibitem[Appel (1985)]{Appel}
 Appel A. 1985, {\it SIAM J. Sci. Stat. Comput.}, 6, 85
\bibitem[Barnes \& Hut (1986)]{BarnesHut}
 Barnes J. \& Hut P. 1986, {\it Nature}   324, 446
\bibitem[Bertschinger \& Gelb(1991)]{GelbBert} Bertschinger E. \& Gelb J. 
 1991, {\it Comp. Phys.}, 5, 164
\bibitem[Bertschinger (1998)]{Bert} Bertschinger, E. 1998, {\it
        Ann. Rev. Astron. Astrophys.,} 36, 599 
\bibitem[Bouchet \& Hernquist(1988) ]{BouchetHernquist} Bouchet F.R.,
	 \& Hernquist L. 1988, \apjs, 68, 521 
\bibitem[Couchman (1991)]{Couchman}  Couchman H.M.P. 1991, \apj, 368, 23
\bibitem[Doroshkevich et al.(1980)]{Doroshkevich} Doroshkevich A.G., Kotok E.V., Novikov I.D.,
	Polyudov A.N., \& Sigov Yu.S. 1980, \mnras, 192, 321
\bibitem[Efstathiou et al. (1985)]{Efff} 
       Efstathiou G., Davis M., Frenk C.S., \& White S.D.M. 1985, \apjs, 57, 241
\bibitem[Gelb (1992)]{Gelb} Gelb J. 1992 {\it Ph.D. Thesis,} MIT
\bibitem[Ghigna et al.(1999)]{Ghigna99} 
	Ghigna, S., Moore, B., Governato, F., Lake, G., Quinn, T., Stadel,
	J. 1999, astro-ph/9910166
\bibitem[Gross (1997)]{Gross} Gross, M. 1997 {\it Ph.D. Thesis,} UC Santa Cruz
\bibitem[Gunn \& Gott (1972)]{GunnGott} Gunn, J.E., \& Gott J.R. 1972, \apj, 176, 1
\bibitem[Hernquist (1987)]{Hernquist} Hernquist L. 1987, \apjs, 64, 715
\bibitem[Hernquist et al. (1991)]{HSB} Hernquist L., Bouchet F.R., \&
 	Suto Y. 1991, \apjs, 75, 231 
\bibitem[Hockney \& Eastwood(1981)]{HockneyEast} 
        Hockney R.W. and Eastwood J.W.  1981, Numerical
        simulations using particles New York: McGraw-Hill
\bibitem[Klypin et al.(1999)]{KGKK} Klypin, A., Gottl\"ober, S., Kravtsov, A., Khokhlov, A. 1999,
\apj, 516, 530 
\bibitem[Klypin \& Shandarin (1983)]{KlypinShandarin}  
	Klypin A., \& Shandarin S.F. 1983, \mnras, 204, 891
\bibitem[Klypin et al.(1993)]{KHPR} Klypin A., Holtzman J., Primack J.,
	\& Regos E. 1993, \apj, 416 1
\bibitem[Klypin \& Holtzman(1997)]{KlypinHoltzman} Klypin A., \&
Holtzman J.1997, astro-ph/9712217
\bibitem[Knebe et al.(1999)]{Knebe}  Knebe, A.,  Kravtsov, A.V.,
Gottl\"ober,  Klypin, A. 1999,  astro-ph/9912257, accpented to MNRAS
\bibitem[Kravtsov et al.(1997)]{ART} Kravtsov A.V., Klypin A.,\&
	Khokhlov A. 1997., Ap. J. Suppl., 111, 73
\bibitem[Kravtsov (1999)]{Kravtsov} Kravtsov, A.V. 1999 {\it Ph.D. Thesis,} New Mexico State University
\bibitem[Peebles (1970)]{Peebles} Peebles P.J.E. 1970, Astron. J.. 75, 13
\bibitem[Press \& Schechter (1974)]{PressSchecter} Press W.H.,
	 \& Schechter P.L. 1974, Ap. J., 187, 425
\bibitem[Quinn et al.(1997)]{Quinn} Quinn, T., Katz, N., \& Stadel, J.,
       Lake, G. 1997, astro=ph/9710043 
\bibitem[Sahni \& Coles(1995)]{SahniColes}  Sahni, V., \& Coles, P. 1995, Physics Reports, 262, 2
\bibitem[Sellwood(1987)]{Sellwood} Sellwood J.A. 1987, {\it
         Ann. Rev. Astron. Astrophys.,} 25, 151 
\bibitem[White (1976)]{WhiteA} White S.D.M. 1976, \mnras, 177, 717
\bibitem[White \& Rees (1978)]{WhiteRees} White S.D.M., \& Rees,
 	M. J. 1978, \mnras, 183, 341
\bibitem[Zeldovich (1970)]{Zeldovich} Zeldovich Ya.B. 1970, \aa, 5, 84

\end{thebibliography}
\end{document}